# Solving Schrödinger equation by mapping it into a Heun-type equation with known solutions

A. D. Alhaidari

*Saudi Center for Theoretical Physics, P. O. Box 32741, Jeddah 21438, Saudi Arabia*

**Abstract**: We transform the Schrödinger wave equation into a nine-parameter Heun-type differential equation. Using our solutions of the latter in [J. Math. Phys. **59** (2018) 113507], we are able to identify the associated potential function, energy parameter, and write down the corresponding wavefunction. Some of the solutions obtained correspond to new integrable quantum systems.

***"To the memory of my friend, colleague and collaborator, the late Mohammed S. Abdelmonem"***

## 1. Introduction

Recently, we introduced the following nine-parameter Heun-type differential equation [1]

$$\left\{\frac{d^2}{dy^2}+\left(\frac{a}{y}-\frac{b}{1-y}-\frac{c}{d-y}\right)\frac{d}{dy}+\left[\frac{A}{y}-\frac{B}{1-y}-\frac{C}{d-y}+yD-E\right]\frac{1}{y(1-y)(d-y)}\right\}\chi(y)=0, \quad (1)$$

where $\{a,b,c,d,A,B,C,D,E\}$ are real dimensionless parameters with $d\neq 0,1$ and positive. We can always take $d>1$. In fact, even if $d<1$ then we can rewrite the equation with the same exact form as (1) but with a redefined set of parameters $\{a',b',c',d',A',B',C',D',E'\}$ such that $d'>1$. This is accomplished by making the replacement $y\to yd$ in (1) resulting in the following redefined parameters

$$d'=d^{-1},\ (a',D')=(a,D),\ (b',c')=(c,b),\ (A',B',C')=(A,C,B)\,d^{-2},\ E'=Ed^{-1}. \quad (2)$$

Thereafter, the solution of Eq. (1) is obtained from the solution of the reparametrized equation as $\chi(y)=\chi'(y/d)$. Now, equation (1) has four regular singularities at $y=\{0,1,d,\infty\}$. The original Heun equation corresponds to $A=B=C=0$ and $D=\alpha\beta$ with the regularity condition at infinity, $\alpha+\beta+1=a+b+c$ [2,3]. If the differential equation parameters $A$ and $B$ are away from their critical values such that $A\le\frac{d}{4}(1-a)^2$ and $B\ge-\frac{d-1}{4}(1-b)^2$ then by using the Tridiagonal Representation Approach (TRA) [4], we were able to obtain four classes of solutions for this equation [1]. The differential equation parameters for each solution class must satisfy the respective constraints shown in Table I. All solutions in the four classes are written as convergent series of square integrable functions $\{\phi_n(y)\}$ as follows

$$\chi(y)=\sum_n f_n\phi_n(y). \quad (3)$$

The expansion coefficients are written as $f_n = f_0 p_n$ and $\{p_n\}$ turn out to be orthogonal polynomials, some of which are either new or modified versions of known ones [1]. The argument and parameters of these polynomials are related to the differential equation parameters. Moreover, the positive definite weight function for $\{p_n\}$ is $f_0^2$. The square integrable basis functions $\{\phi_n(y)\}$ are written in terms of the Jacobi polynomial $P_n^{(\mu,\nu)}(y)$ as follows

$$\phi_n(y) = \mathcal{A}_n y^\alpha (1-y)^\beta (d-y)^\gamma P_n^{(\mu,\nu)}(y), \tag{4}$$

where the normalization constant is conveniently chosen as $\mathcal{A}_n = \sqrt{(2n+\mu+\nu+1)\frac{\Gamma(n+1)\Gamma(n+\mu+\nu+1)}{\Gamma(n+\mu+1)\Gamma(n+\nu+1)}}$. The basis parameters $\{\alpha,\beta,\gamma,\mu,\nu\}$ are related to the differential equation parameters as shown in Table II for each of the four solution classes. The Jacobi polynomial is defined as

$$P_n^{(\mu,\nu)}(y) = \frac{\Gamma(n+\mu+1)}{\Gamma(n+1)\Gamma(\mu+1)}\, {}_2F_1\left(\begin{matrix}-n, n+\mu+\nu+1\\ \mu+1\end{matrix}\middle| 1-y\right), \tag{5}$$

with $\mu > -1$, $\nu > -1$ and $y \in [0,+1]$. This polynomial definition is obtained by the replacement $y \to 2y-1$ in the classical definition in which $y \in [-1,+1]$.

In this work, we make a transformation that maps the Schrödinger wave equation into Eq. (1). Consequently, we will be able to identify the potential function, energy parameter and corresponding solutions (wavefunctions) using the results obtained in [1]. The transformation employed is a combination of independent and dependent variable transformation (i.e., coordinate and wavefunction transformations). In the atomic units $\hbar = m = 1$, the one-dimensional time-independent Schrödinger equation reads as follows

$$\left[-\frac{1}{2}\frac{d^2}{dx^2} + V(x) - \mathcal{E}\right]\psi(x) = 0, \tag{6}$$

where $V(x)$ is the potential function and $\mathcal{E}$ is the energy. In three dimensions with spherical symmetry, this equation could also be taken as the radial Schrödinger equation with $x \mapsto r$ and $V(x) \mapsto \frac{\ell(\ell+1)}{2r^2} + V(r)$, where $\ell$ is the angular momentum quantum number and $V(r)$ is the radial interaction potential. Now, we make the coordinate transformation $x \mapsto y(x)$ in Eq. (6) and write $\psi(x) = h(y)\chi(y)$. If we define $\frac{dy}{dx} = \lambda g(y)$, where $\lambda$ is a positive scale parameter with inverse length dimension, then these transformations map Eq. (6) into the following second order differential equation in the dimensionless variable $y$

$$hg^2\left[\frac{d^2}{dy^2} + \left(\frac{g'}{g} + 2\frac{h'}{h}\right)\frac{d}{dy} + \frac{h''}{h} + \frac{g'h'}{gh} - \frac{U}{g^2}\right]\chi(y) = 0, \tag{7}$$

where $U = \frac{2}{\lambda^2}(V - \mathcal{E})$ and the prime stands for the derivative with respect to $y$. Identifying this equation with (1) and assuming that $hg^2 \neq 0$ within the open interval $y \in (0,+1)$ dictate that

$$\frac{g'}{g}+2\frac{h'}{h}=\frac{a}{y}-\frac{b}{1-y}-\frac{c}{d-y}, \tag{8a}$$

$$U=\frac{g^2}{y(1-y)(d-y)}\left[-\frac{A}{y}+\frac{B}{1-y}+\frac{C}{d-y}-yD+E+y(1-y)(d-y)\left(\frac{h''}{h}+\frac{g'h'}{gh}\right)\right]. \tag{8b}$$

Compatibility of the wavefunction expansion (3) and basis ansatz (4) with Eq. (8a) suggest that we take $g(y)=y^a(1-y)^b$ and $h(y)=(d-y)^{c/2}$. Thus, the configuration space coordinate could be obtained in terms of $y$ by evaluating the integral $\lambda x=\int y^{-a}(1-y)^{-b}dy$. This is done in Appendix A where $x(y)$ is written in terms of the incomplete beta functions [5,6]. The inverse transform that gives the coordinate transformation $y(x)$ is simple only for special values of the parameters $a$ and $b$. Now, substituting $g(y)$ and $h(y)$ in Eq. (8b) gives

$$\frac{2}{\lambda^2}(V-\mathcal{E})=\frac{y^{2a-1}(1-y)^{2b-1}}{d-y}\left[E+\tilde{E}+y\left(\tilde{D}-D\right)+\frac{C+\tilde{C}}{d-y}-\frac{A}{y}+\frac{B}{1-y}\right], \tag{9}$$

where $\tilde{C}=-\frac{1}{4}cd(c-2)(d-1)$, $\tilde{D}=\frac{1}{4}c(2a+2b+c-2)$, and $\tilde{E}=\frac{1}{4}c\left[(c-2)(d-1)-2a\right]$. Moreover, the wavefunction series becomes

$$\psi(x)=h(y)\chi(y)=f_0(z)y^{\alpha}(1-y)^{\beta}(d-y)^{\gamma+c/2}\sum_n \mathcal{A}_n p_n(z)P_n^{(\mu,\nu)}(y), \tag{10}$$

where $z$ is some proper function of the differential equation parameters. In the ensuing analysis, we consider potential functions that satisfy the following two conditions:

(1) They are energy independent functions, and
(2) They vanish at infinity (i.e., at $+\infty$ if $x\ge 0$, and at $-\infty$ or at $\pm\infty$ if $-\infty<x<+\infty$).

Thus, the parameters $a$ and $b$ must be chosen such that the right-hand side of Eq. (9) contains a constant to be identified with the energy parameter $-2\mathcal{E}/\lambda^2$ on the left. We observe that the individual terms inside the square brackets in Eq. (9) are linear polynomials in $y$ raised to the power zero or $\pm 1$. Therefore, a necessary (but not sufficient) condition for the right-hand side of Eq. (9) to contain a constant is that the factor $y^{2a-1}(1-y)^{2b-1}(d-y)^{-1}$ becomes a ratio of two polynomials in $y$ each with a maximum degree of two but such that the difference between their two degrees is either zero or $\pm 1$. Consequently, it is necessary, but not sufficient, that the parameters $a$ and $b$ assume only half integer values and must belong to the set $\left\{0,\frac{1}{2},1,\frac{3}{2}\right\}$. However, the stated constraint on the degrees of the polynomial ratios in the factor $y^{2a-1}$ $(1-y)^{2b-1}(d-y)^{-1}$ leaves only ten out of the possible sixteen choices. Moreover, due to the exchange symmetry $a\leftrightarrow b$ and $y\leftrightarrow 1-y$ we end up with only six independent choices for the parameter set $\{a,b\}$ out of the possible ten. For each of these six choices, we give in Table III the corresponding coordinate transformation, potential function and energy parameter. The constants in curly brackets have been added by hand to the potentials to force them to vanish at infinity. Of course, the same constants were added to the energy. The two cases in the last two rows of Table III corresponding to $b=\frac{3}{2}$ are solvable only at zero energy. Obviously, if we allow some of the parameters in these potentials to depend on the energy, then we could also obtain non-zero energy solutions. This will not be done in this work, but a brief discussion of this issue will be presented at the end of the Conclusion section with an example. In all subsequent treatment, these two zero energy solutions will be ignored. Note that although the energy parameter appears in the potential function, a simple parameter redefinition eliminates this superficial energy dependence of the potential. For example, in the case $(a,b)=\left(\frac{1}{2},1\right)$ the parameter redefinition $E\mapsto E+B/(d-1)$ will eliminate the energy parameter $B/(1-d)$ from the potential function. It is interesting to note that the energy parameter for the potential box corresponding to $(a,b)=\left(\frac{1}{2},\frac{1}{2}\right)$ is $\tilde{D}-D$, which is equal to $\frac{1}{4}c^2-D$. On the other hand, for all other cases with unbounded configuration space, it is the same parameter, $B/(1-d)$.

Since Table II gives the basis parameters $\{\alpha,\beta,\gamma,\mu,\nu\}$ in terms of the differential equation parameters $\{a,b,c,A,B\}$ then the basis elements (4) are completely determined. Hence, for a full realization of the solution of the Schrödinger equation as given by the series expansion (10), we only need to obtain the polynomial coefficients $\{p_n(z)\}$ and their corresponding weight function $f_0^2(z)$. In the following three sections, we do that for all solution classes derived in [1]. It is shown elsewhere that the physical properties of any system in a given class is determined from the properties of the corresponding orthogonal polynomial (e.g., its weight function, generating function, asymptotics, spectrum formula, zeros, etc.) [4,7]. However, these properties are known only for the Wilson polynomial, which is associated with the restricted solution class. The other two solution classes are associated either with a new polynomial or with a modified version of the Wilson polynomial. The properties of both are not known yet in the published literature. This remains an open problem in orthogonal polynomials. For a presentation of this and other related open problems, the reader may consult Ref. [8]. Nonetheless, these polynomials could be written explicitly to all degrees (albeit not in a closed form) using their respective recursion relation and initial value $p_0(z)=1$. Consequently, the series representation of the wavefunction (10) is fully determined and, in accordance with the postulates of quantum mechanics, the corresponding physical system is well defined. In the rest of the paper, we identify the orthogonal polynomials in the solution series for each class then end with a conclusion and some remarks.

## 2. The general solution class

Table IV is a reproduction of Table III after imposing the class constraint $4C=(1-c)^2d(d-1)$ , which makes $C+\tilde{C}=d(d-1)/4$. Aside from the scale parameter $\lambda$, Table IV shows that all potentials in this class have four parameters. Now, the orthogonal polynomial associated with this class of solutions satisfy the symmetric three-term recursion relation obtained in Appendix B of Ref. [1] as Eq. (B10b), which reads

$$\begin{aligned}R\,p_n(z)&=G_{n-1}\left(S_{n-1}+D\right)p_{n-1}(z)+G_n\left(S_n+D\right)p_{n+1}(z)\\&\left[-\frac{n(n+\mu)}{2n+\mu+\nu}+d\left(n+\tfrac{\mu+\nu+1}{2}\right)+\frac{1}{2}\left(F_n+1-2d\right)\left(S_n+D\right)-\frac{1}{4}\left(\nu+1\right)^2\right]p_n(z)\end{aligned}\tag{11}$$

for $n=0,1,2,...$ with $p_0(z)=1$, $p_{-1}(z):=0$ and where

$$R = E - \frac{B}{d-1} - Dd - \frac{1}{4}(1-a)^2 + \frac{c}{2}\left[d\left(a+b+c-2\right) - a - \frac{c}{2} + 1\right], \tag{12a}$$

$$S_n = \left(n + \tfrac{\mu+\nu}{2} + 1\right)^2 - \frac{1}{4}\left(a+b+c-1\right)^2, \quad F_n = \frac{\nu^2 - \mu^2}{\left(2n+\mu+\nu\right)\left(2n+\mu+\nu+2\right)}, \tag{12b}$$

$$G_n = \frac{1}{2n+\mu+\nu+2}\sqrt{\frac{\left(n+1\right)\left(n+\mu+1\right)\left(n+\nu+1\right)\left(n+\mu+\nu+1\right)}{\left(2n+\mu+\nu+1\right)\left(2n+\mu+\nu+3\right)}}. \tag{12c}$$

Comparing (11) to the recursion relation of the "Racah-Heun polynomial" given in Appendix B by (B5), we conclude that $p_n(z) = \left(-1\right)^n W_n^\kappa(z^2;\sigma-\tau,\sigma+\tau,\eta,\eta)$ where

$$\kappa = d, \quad 2\sigma = \mu+1, \quad 2\eta = \nu+1, \quad 2\tau = \sqrt{\left(a+b+c-1\right)^2 - 4D}, \quad z^2 = R. \tag{13}$$

In the comparison, we have used the following identity:

$$\frac{\left(n+\nu+1\right)\left(n+\mu+\nu+1\right)\left(S_n+D\right)}{\left(2n+\mu+\nu+1\right)\left(2n+\mu+\nu+2\right)} + \frac{n\left(n+\mu\right)\left(S_{n-1}+D\right)}{\left(2n+\mu+\nu\right)\left(2n+\mu+\nu+1\right)} = -\frac{n\left(n+\mu\right)}{2n+\mu+\nu} + \frac{1}{2}(1+F_n)(S_n+D). \tag{14}$$

Finally, the exact series representation of the wavefunction in this general class is written as:

$$\psi(x) = \sqrt{\rho^\kappa(z) y^{\nu+1-a}(1-y)^{\mu+1-b}(d-y)} \sum_n \left(-1\right)^n \mathcal{A}_n W_n^\kappa(z^2;\sigma-\tau,\sigma+\tau,\eta,\eta) P_n^{(\mu,\nu)}(y), \tag{15}$$

where $\rho^\kappa(z)$ is the weight function for $W_n^\kappa(z^2;\sigma-\tau,\sigma+\tau,\eta,\eta)$.

## 3. The special solution class

Table V is a reproduction of Table III after imposing the class constraint $4C = c(c-2)d(d-1)$ , which makes $C + \tilde{C} = 0$. In addition to the scale parameter $\lambda$, Table V shows that all potential functions in this class have four parameters. The orthogonal polynomials associated with this class of solutions satisfy the symmetric three-term recursion relation obtained in Appendix B of Ref. [1] as Eq. (B10a) that reads

$$\left(2d-1\right) p_n(z) = \left[\frac{-2\tilde{R}}{T_n + D} + F_n\right] p_n(z) + 2G_{n-1} p_{n-1}(z) + 2G_n p_{n+1}(z), \tag{16}$$

for $n = 0,1,2,...$ with $p_0(z) = 1$, $p_{-1}(z) := 0$ and where

$$\tilde{R} = R - \frac{A}{d} + \frac{1}{4}\left(1-a\right)^2, \qquad T_n = \left(n + \tfrac{\mu+\nu+1}{2}\right)^2 - \frac{1}{4}\left(a+b+c-1\right)^2 = S_{n-\frac{1}{2}} \tag{17}$$

Comparing (16) to the recursion relation of the recently introduced orthogonal polynomial $V_n^{(\mu,\nu)}(z;\tau,\theta)$, which is given in Appendix C by (C1), we conclude that $p_n(z) = V_n^{(\mu,\nu)}(z;\tau,\theta)$ where

$$\cosh\theta = 2d-1,\quad 2\tau = \sqrt{(a+b+c-1)^2-4D}\,,\quad z = -\tilde{R}\big/\sqrt{d(d-1)}\,. \tag{18}$$

Therefore, the wavefunction associated with this special class of solutions is written as the following series:

$$\psi(x) = \sqrt{\omega(z)y^{\nu+1-a}(1-y)^{\mu+1-b}}\sum_n \mathcal{A}_n V_n^{(\mu,\nu)}(z;\tau,\theta)P_n^{(\mu,\nu)}(y)\,. \tag{19}$$

where $\omega(z)$ is the weight function for $V_n^{(\mu,\nu)}(z;\tau,\theta)$.

# 4. The restricted solution class

Due to the exchange symmetry between the two families of solution in this class, we consider only the first family corresponding to the third column of Table II. Table VI is a reproduction of Table III after imposing the class constraints shown in the third column of Table I. That is,

$$D=0\,,\qquad C+\tilde{C}=0\,,\qquad E+\tilde{E}=\frac{A}{d}+\frac{B}{d-1}-\tilde{D}d\,, \tag{20}$$

Aside from the scale parameter $\lambda$, the number of potential parameters are three except for the case corresponding to $(a,b)=(1,1)$ where it is two. Note that the parameter relation in Table VI for the case $(a,b)=(0,1)$ imply that the parameter $c$ in this case is energy dependent but such that the quantity $4B-(d-1)c^2$ dos not depend on the energy. As an explicit illustration for this class, we list in Table VII the potential functions in the configuration space coordinate showing the relations of the potential parameters to the differential equation parameters. The orthogonal polynomials associated with this restricted class of solutions satisfy the following symmetric three-term recursion relation found in Appendix B of Ref. [1] as Eq. (B10c), which reads

$$\begin{aligned}\frac{1}{4}(\mu+1)^2 p_n(z) = \left[\frac{n(n+\nu)}{2n+\mu+\nu}+\frac{1}{2}(F_n-1)S_n+\frac{1}{4}(\mu+1)^2\right]p_n(z)\\ +G_{n-1}S_{n-1}p_{n-1}(z)+G_nS_np_{n+1}(z)\end{aligned} \tag{21}$$

for $n=0,1,2,...$ and with $p_0(z)=1$, $p_{-1}(z):=0$. Comparing this to the recursion relation of the normalized version of the Wilson polynomial given in Appendix B by Eq. (B2), we conclude that $p_n(z)=W_n(z^2;\sigma-\tau,\sigma+\tau,\eta,\eta)$ where

$$2\sigma=\nu+1,\qquad 2\eta=\mu+1,\qquad 2\tau=a+b+c-1,\qquad 4z^2=-(\mu+1)^2\,. \tag{22}$$

In the comparison, we used identity (14) after making the exchange $\mu\leftrightarrow\nu$. Since $z^2<0$, then the spectrum is purely discrete and the spectrum formula (B4c) gives

$$\frac{2\mathcal{E}}{\lambda^2}=\frac{-B}{d-1}=-\left(k+1+\tfrac{\nu-a-b-c}{2}\right)^2+\frac{1}{4}(1-b)^2, \tag{23}$$

where $k=0,1,2,...N$ and $N$ is the largest integer less than or equal to $\tau-\sigma=\frac{a+b+c-\nu}{2}-1$. The parameter values in Table VI and Table II show that the size of the energy spectrum $N$ depends on the parameters $c$ and $A/d$. The larger the positive value of $c$ and the closer $A/d$ to its critical value $\frac{1}{4}(1-a)^2$, the larger the value of $N$. One may test the validity of the spectrum formula (23) by comparing with well-known results, for example, with the energy spectrum of the Scarf potential shown in the first row of Table VII or the Pöschl-Teller potential in the second row. Finally, the wavefunction associated with this restricted class of solutions is written as follows:

$$\psi(x)=\sqrt{\rho(k)y^{\nu+1-a}(1-y)^{\mu+2-b}}\sum_n \mathcal{A}_n W_n(z^2;\sigma-\tau,\sigma+\tau,\eta,\eta)P_n^{(\mu,\nu)}(y), \tag{24}$$

where $\rho(z)$ is the weight function of the Wilson polynomial $W_n(z^2;\sigma-\tau,\sigma+\tau,\eta,\eta)$.

# 5. Conclusion

We made a combined coordinate and wavefunction transformation of the one-dimensional Schrödinger equation. The transformed equation is identified with the nine-parameter Heun-type equation that we have already solved in an earlier publication [1]. Consequently, we were able to identify the potential function and energy parameter associated with any given solution class of the Heun-type equation found in [1]. Moreover, the series solution of the equation obtained in our earlier publication for each class is identified with the wavefunction that solves the original Schrödinger equation.

In closing, we add some relevant comments. Firstly, it is remarkable to see the numerous novel and exactly solvable potentials associated with the Heun equation (1). Table III is a testimony to the richness of this class of potentials. Secondly, even the two energy-dependent potentials that we have ignored are worthy of further investigation. For example, the case that corresponds to $(a,b)=\left(\frac{1}{2},\frac{3}{2}\right)$ in the restricted solution class leads to the following potential

$$V(r)=\frac{1}{\left(1+\frac{1}{4}\lambda^2r^2\right)^2}\left[\frac{\ell(\ell+1)}{2r^2}+\frac{1}{2}K^2r^2+\omega_0^2\right], \tag{26}$$

where $K$ is an energy dependent parameter and $\omega_0$ is the system's natural frequency. The parameters $\{\ell,k,\omega_0\}$ can easily be written in terms of the differential equation parameters. This potential is an interesting deformation of the isotropic oscillator potential with $\lambda$ being the deformation parameter. For a given range of these parameters (i.e., an energy range), the potential can support resonances as well as bound states. In fact, since it vanishes at infinity it can also have continuum scattering states.

## Appendix A: Coordinate transformation

The coordinate transformation $x \to y(x)$ is such that $0 \le y(x) \le 1$ and $\frac{dy}{dx} = \lambda y^a (1-y)^b$. Integration yields $\lambda x = \int y^{-a}(1-y)^{-b} dy$, which is a match with the integral representation of the incomplete beta function [5,6]. Thus, we obtain two results depending on the physical space (values of the parameter *a* and *b*). The first is

$$\int_0^y t^{-a}(1-t)^{-b} dt = B\left(y;1-a,1-b\right), \tag{A1}$$

where $B(z;\alpha,\beta)$ is the lower incomplete beta function, which is equal to $\left(z^\alpha/\alpha\right)\,{}_2F_1\left(\begin{smallmatrix}\alpha,1-\beta\\1+\alpha\end{smallmatrix}\middle|z\right)$. Therefore, we obtain (for $a \ne 1$)

$$\lambda x = \int_0^y t^{-a}(1-t)^{-b} dt = B\left(y;1-a,1-b\right) = \frac{y^{1-a}}{1-a}\,{}_2F_1\left(\begin{smallmatrix}b,1-a\\2-a\end{smallmatrix}\middle|y\right), \tag{A2}$$

On the other hand, changing the integration limits and after some simple manipulations, we obtain the alternative result (the upper incomplete beta function)

$$\int_y^1 t^{-a}(1-t)^{-b} dt = B\left(1-y;1-b,1-a\right), \tag{A3}$$

where $B\left(y;\alpha,\beta\right) + B\left(1-y;\beta,\alpha\right) = B(\alpha,\beta) = \frac{\Gamma(\alpha)\Gamma(\beta)}{\Gamma(\alpha+\beta)}$ is the complete beta function. Therefore, we obtain the alternative result (for $b \ne 1$)

$$\lambda x = \int_y^1 t^{-a}(1-t)^{-b} dt = B\left(1-y;1-b,1-a\right) = \frac{(1-y)^{1-b}}{1-b}\,{}_2F_1\left(\begin{smallmatrix}a,1-b\\2-b\end{smallmatrix}\middle|1-y\right). \tag{A4}$$

For the singular case $a = b = 1$ direct integration gives $y(x) = \frac{1}{1+e^{-\lambda x}}$ for $-\infty < x < +\infty$ corresponding $0 \le y \le 1$.

## Appendix B: The Wilson and Racah-Heun polynomials

Some symbols in this Appendix are local and not related to those in the rest of the paper. According to Favard's theorem [9,10], a polynomial sequence $\{P_n(x)\}$ that satisfies the three-term recursion relation $xP_n(x) = A_n P_n(x) + B_{n-1}P_{n-1}(x) + C_n P_{n+1}(x)$ is orthogonal with respect to a positive definite weight function if $B_n C_n > 0$ for all *n*. The normalized version of this polynomial satisfies a symmetric three-term recursion relation of the form $xp_n(x) = a_n p_n(x) + b_{n-1}p_{n-1}(x) + b_n p_{n+1}(x)$ with $b_n^2 = B_n C_n$, $a_n = A_n$ and $p_0(x) = 1$. The completely continuous version of this polynomial has an orthogonality that reads: $\int_{x_-}^{x_+} \rho(x) p_n(x) p_m(x) dx = \delta_{n,m}$ where $\rho(x)$ is the positive definite normalized weight function. However, if the spectrum is a mix of continuous and discrete parts then this orthogonality is modified by the addition of a discrete (finite or infinite) sum. Now, the normalized version of the four-parameter Wilson polynomial is written as (see, Appendix A in Ref. [11])

$$W_n(z^2;a,b,c,d)=\sqrt{\left(\frac{2n+a+b+c+d-1}{n+a+b+c+d-1}\right)\frac{(a+b)_n(a+c)_n(a+d)_n(a+b+c+d)_n}{(b+c)_n(b+d)_n(c+d)_n\,n!}}\;{}_4F_3\left(\left.\begin{matrix}-n,n+a+b+c+d-1,a+\mathrm{i}z,a-\mathrm{i}z\\ a+b,a+c,a+d\end{matrix}\right|1\right) \tag{B1}$$

It satisfies the following symmetric three-term recursion relation

$$\begin{aligned}z^2W_n=&\left[\frac{(n+a+b)(n+a+c)(n+a+d)(n+a+b+c+d-1)}{(2n+a+b+c+d)(2n+a+b+c+d-1)}+\frac{n(n+b+c-1)(n+b+d-1)(n+c+d-1)}{(2n+a+b+c+d-1)(2n+a+b+c+d-2)}-a^2\right]W_n\\&-\frac{1}{2n+a+b+c+d-2}\sqrt{\frac{n(n+a+b-1)(n+c+d-1)(n+a+c-1)(n+a+d-1)(n+b+c-1)(n+b+d-1)(n+a+b+c+d-2)}{(2n+a+b+c+d-3)(2n+a+b+c+d-1)}}\,W_{n-1}\\&-\frac{1}{2n+a+b+c+d}\sqrt{\frac{(n+1)(n+a+b)(n+c+d)(n+a+c)(n+a+d)(n+b+c)(n+b+d)(n+a+b+c+d-1)}{(2n+a+b+c+d-1)(2n+a+b+c+d+1)}}\,W_{n+1}\end{aligned} \tag{B2}$$

If $\mathrm{Re}(a,b,c,d)>0$ and non-real parameters occur in conjugate pairs, then the polynomial has only a continuous positive spectrum with the following normalized weight function

$$\rho(z)=\frac{1}{2\pi}\frac{\Gamma(a+b+c+d)\left|\Gamma(a+\mathrm{i}z)\Gamma(b+\mathrm{i}z)\Gamma(c+\mathrm{i}z)\Gamma(d+\mathrm{i}z)/\Gamma(2\mathrm{i}z)\right|^2}{\Gamma(a+b)\Gamma(c+d)\Gamma(a+c)\Gamma(a+d)\Gamma(b+c)\Gamma(b+d)}. \tag{B3}$$

On the other hand, if the parameters are such that $a<0$ and $a+b$, $a+c$, $a+d$ are positive or a pair of complex conjugates with positive real parts, then the polynomial will have a mix of continuous positive spectrum and a finite-size negative discrete spectrum and the polynomial satisfies a generalized orthogonality relation [see, Eq. (C3) in Ref. 7 and Eq. (9.1.3) in Ref. 12]. The asymptotics ($n\to\infty$) of the Wilson polynomial gives the following scattering amplitude, phase shift and spectrum formula, respectively (see Appendix B in Ref. [11])

$$A(z)=2\Big/\sqrt{\pi\,\rho(z)}, \tag{B4a}$$

$$\delta(z)=\arg\Gamma(2\mathrm{i}z)-\arg\left[\Gamma(a+\mathrm{i}z)\Gamma(b+\mathrm{i}z)\Gamma(c+\mathrm{i}z)\Gamma(d+\mathrm{i}z)\right], \tag{B4b}$$

$$z_k^2=-(k+a)^2, \tag{B4c}$$

where $k=0,1,..,N$ and $N$ is the largest integer less than or equal to $-a$.

The modified version of the Wilson polynomial (named the "Racah-Heun polynomial" by the authors of Ref. [13]) is written as $W_n^\lambda(z^2;a,b,c,d)$, where $\lambda$ is the deformation (modification) parameter. It satisfies a modified version of the three-term recursion relation (B2) that reads

$$\begin{aligned}z^2W_n^\lambda&=A_nW_n^\lambda-B_{n-1}W_n^\lambda-B_nW_{n+1}^\lambda\\&-\lambda\left[(n+a+c)(n+b+d)-\tfrac{1}{2}(2n+a+b+c+d-1)\right]W_n^\lambda\end{aligned} \tag{B5}$$

where $\{A_n,B_n\}$ are the recursion coefficients in (B2). All properties of the polynomial $W_n^\lambda(z^2;a,b,c,d)$ are yet to be derived analytically. This is still an open problem in orthogonal polynomials [8].

# Appendix C: The new orthogonal polynomial associated with the class of special solutions

This polynomial, referred to as $V_n^{(\mu,\nu)}(z;\tau,\theta)$, was introduced as an open problem in orthogonal polynomials [8]. It is defined, up to now, by its three-term recursion relation and initial value $V_0^{(\mu,\nu)}(z;\tau,\theta)=1$. Its other properties (weight function, generating function, orthogonality, asymptotics, zeros, etc.) are yet to be derived analytically. Its normalized version satisfies the following symmetric three-term recursion relation

$$\begin{aligned}(\cosh\theta)V_n^{(\mu,\nu)}(z;\tau,\theta)=\left\{z(\sinh\theta)\left[\left(n+\tfrac{\mu+\nu+1}{2}\right)^2-\tau^2\right]^{-1}+F_n\right\}V_n^{(\mu,\nu)}(z;\tau,\theta)\\ +2G_{n-1}V_{n-1}^{(\mu,\nu)}(z;\tau,\theta)+2G_nV_{n+1}^{(\mu,\nu)}(z;\tau,\theta)\end{aligned} \tag{C1}$$

where $\theta\geq 0$ and $\{F_n,G_n\}$ are defined by Eq. (12) in the text above. It was conjectured that if $\tau$ is pure imaginary, then the spectrum is purely continuous and positive. However, if $\tau$ is real then the spectrum is a mix of a continuous positive spectrum and a discrete negative spectrum of finite size $N+1$, where $N$ is the largest integer less than or equal to $|\tau|-\frac{\mu+\nu+1}{2}$. In that case, the polynomial satisfies the following generalized orthogonality

$$\int_0^\infty \rho(z)V_n^{(\mu,\nu)}(z;\tau,\theta)V_m^{(\mu,\nu)}(z;\tau,\theta)\,dz+\sum_{k=0}^{N}\omega(k)V_n^{(\mu,\nu)}(z_k;\tau,\theta)V_m^{(\mu,\nu)}(z_k;\tau,\theta)=\delta_{n,m}, \tag{C2}$$

where $\rho(z)$ and $\omega(k)$ are the positive definite continuous and discrete weight functions, respectively. The finite discrete spectrum $\{z_k\}_{k=0}^{N}$ could be determined from the condition that forces the asymptotics ($n\to\infty$) of $V_n^{(\mu,\nu)}(z;\tau,\theta)$ to vanish. It remains an open problem to determine the discrete spectrum and weight functions analytically.

# Tables Caption:

**Table I**: The conditions on the parameters of the differential equation (1) in each of its four solution classes.

**Table II**: The parameters $\{\alpha,\beta,\gamma,\mu,\nu\}$ of the basis elements (4) in terms of the differential equation parameters for each of the four solution classes.

**Table III**: The six choices for the parameter set $(a,b)$ along with the corresponding coordinate transformation, energy parameter and potential function. Parameters redefinition shows how to eliminate the superficial energy dependence in the corresponding potential.

**Table IV**: The potential function and energy parameter for each of the four cases corresponding to a given set of parameters $(a,b)$ in the class of general solutions.

**Table V**: The potential function and energy parameter for each of the four cases corresponding to a given set of parameters $(a,b)$ in the class of special solutions.

**Table VI**: The potential function and energy parameter for each of the three cases corresponding to a given set of parameters $(a,b)$ in the class of restricted solutions.

**Table VII**: The potential functions of Table VI written explicitly in the configuration space coordinate *x*.

**Table I**

| General Solution | Special Solution | Two Restricted Solutions |
|---|---|---|
| $4\frac{A}{d} \le (1-a)^2$ | $4\frac{A}{d} \le (1-a)^2$ | $4\frac{A}{d} \le (1-a)^2$ |
| $\frac{4B}{d-1} \ge -(1-b)^2$ | $\frac{4B}{d-1} \ge -(1-b)^2$ | $\frac{4B}{d-1} \ge -(1-b)^2$ |
| $\frac{4C}{d(d-1)} = (1-c)^2$ | $\frac{4C}{d(d-1)} = (1-c)^2 - 1$ | $\frac{4C}{d(d-1)} = (1-c)^2 - 1$ |
| | | $D = 0$ |
| | | $E = \frac{A}{d} + \frac{B}{d-1} - \frac{c}{2}\left[d\left(a+b+c-2\right) - a - \frac{c}{2} + 1\right]$ |

**Table II**

| General Solution | Special Solution | First Restricted Solution | Second Restricted Solution |
|---|---|---|---|
| $2\alpha = \nu + 1 - a$ | $2\alpha = \nu + 1 - a$ | $2\alpha = \nu + 1 - a$ | $2\alpha = \nu + 2 - a$ |
| $2\beta = \mu + 1 - b$ | $2\beta = \mu + 1 - b$ | $2\beta = \mu + 2 - b$ | $2\beta = \mu + 1 - b$ |
| $2\gamma = 1 - c$ | $2\gamma = -c$ | $2\gamma = -c$ | $2\gamma = -c$ |
| $\nu^2 = (1-a)^2 - 4\frac{A}{d}$ | $\nu^2 = (1-a)^2 - 4\frac{A}{d}$ | $\nu^2 = (1-a)^2 - 4\frac{A}{d}$ | $(\nu+1)^2 = (1-a)^2 - 4\frac{A}{d}$ |
| $\mu^2 = (1-b)^2 + \frac{4B}{d-1}$ | $\mu^2 = (1-b)^2 + \frac{4B}{d-1}$ | $(\mu+1)^2 = (1-b)^2 + \frac{4B}{d-1}$ | $\mu^2 = (1-b)^2 + \frac{4B}{d-1}$ |

**Table III**

| $(a,b)$ | $y(x)$ | $2\mathcal{E}/\lambda^2$ | $2V(x)/\lambda^2$ | Parameter Redefinition |
|---|---|---|---|---|
| $\left(\frac{1}{2},\frac{1}{2}\right)$ | $\frac{1}{2}[1+\sin(\lambda x)]$, $-\frac{\pi}{2}\le\lambda x\le+\frac{\pi}{2}$ | $\frac{1}{4}c^2-D$ | $\frac{1}{d-y}\left[E+\tilde{E}+d\left(\frac{1}{4}c^2-D\right)-\frac{A}{y}+\frac{B}{1-y}+\frac{C+\tilde{C}}{d-y}\right]$ | $E\mapsto E+d\left(D-\frac{1}{4}c^2\right)$ |
| $\left(\frac{1}{2},1\right)$ | $\tanh^2(\lambda x/2)$, $x\ge 0$ | $\frac{-B}{d-1}$ | $\frac{1-y}{d-y}\left[y\left(\frac{c}{4}(c+1)-D\right)-\frac{A}{y}+\frac{C+\tilde{C}}{d-y}\right]+\frac{B-(d-1)(E+\tilde{E})}{d-y}+\left\{E+\tilde{E}-\frac{B}{d-1}\right\}$ | $E\mapsto E+\frac{B}{d-1}$ |
| $(1,1)$ | $(1+e^{-\lambda x})^{-1}$, $-\infty\le\lambda x\le+\infty$ | $\frac{-B}{d-1}$ | $\frac{y(1-y)}{d-y}\left[E+\tilde{E}+y\left(\frac{c}{4}(c+2)-D\right)+\frac{C+\tilde{C}}{d-y}\right]+\frac{(d-1)A+Bd}{d-y}+\left\{-A-\frac{Bd}{d-1}\right\}$ | $A\mapsto A-\frac{Bd}{d-1}$ |
| $(0,1)$ | $1-e^{-\lambda x}$, $x\ge 0$ | $\frac{-B}{d-1}$ | $\frac{1-y}{y(d-y)}\left[-\frac{A}{y}+\frac{C+\tilde{C}}{d-y}\right]+\frac{1}{d}\frac{E+\tilde{E}+B}{y}+\left(\frac{d-1}{d}\right)\frac{\left(D-\frac{1}{4}c^2\right)d-E-\tilde{E}+B/(d-1)}{d-y}+\left\{\frac{1}{4}c^2-D-\frac{B}{d-1}\right\}$ | $D\mapsto D-\frac{B}{d-1}$ <br> $E\mapsto E-B$ |
| $\left(\frac{1}{2},\frac{3}{2}\right)$ | $\frac{(\lambda x/2)^2}{(\lambda x/2)^2+1}$, $x\ge 0$ | 0 | $\frac{(1-y)^2}{d-y}\left[E+\tilde{E}+y\left(\tilde{D}-D\right)+\frac{C+\tilde{C}}{d-y}-\frac{A}{y}+\frac{B}{1-y}\right]$ | None |
| $\left(0,\frac{3}{2}\right)$ | $1-\frac{1}{(\lambda x/2)^2}$, $x\ge 2\lambda^{-1}$ | 0 | $\frac{(1-y)^2}{d-y}\left[E+\tilde{E}+y\left(\tilde{D}-D\right)+\frac{C+\tilde{C}}{d-y}-\frac{A}{y}+\frac{B}{1-y}\right]$ | None |

**Table IV**

| $(a,b)$ | $y(x)$ | $2\mathcal{E}/\lambda^2$ | $2V(x)/\lambda^2$ $[u_i = 2V_i/\lambda^2]$ | Parameter Relations |
|---|---|---|---|---|
| $\left(\frac{1}{2},\frac{1}{2}\right)$ | $\frac{1}{2}[1+\sin(\lambda x)]$, $-\frac{\pi}{2} \le \lambda x \le +\frac{\pi}{2}$ | $\frac{1}{4}c^2 - D$ | $\frac{1}{d-y}\left[u_0 - \frac{A}{y} + \frac{B}{1-y} + \frac{d(d-1)/4}{d-y}\right]$ | $E = u_0 - \tilde{E} + d(D - \frac{1}{4}c^2)$ |
| $\left(\frac{1}{2},1\right)$ | $\tanh^2(\lambda x/2)$, $x \ge 0$ | $\frac{-B}{d-1}$ | $\frac{1-y}{d-y}\left[u_1 y + \frac{u_0}{1-y} - \frac{A}{y} + \frac{d(d-1)/4}{d-y}\right] - \frac{u_0}{d-1}$ | $E = \frac{B-u_0}{d-1} - \tilde{E}$ <br> $D = \frac{1}{4}c(c+1) - u_1$ |
| $(1,1)$ | $(1+e^{-\lambda x})^{-1}$, $-\infty \le \lambda x \le +\infty$ | $\frac{-B}{d-1}$ | $\frac{y(1-y)}{d-y}\left[u_2 + u_1 y + \frac{d(d-1)/4}{d-y}\right] + \frac{u_0}{d-y} - \frac{u_0}{d-1}$ | $A = \frac{u_0 - Bd}{d-1}$ <br> $D = \frac{1}{4}c(c+2) - u_1$ <br> $E = u_2 - \tilde{E}$ |
| $(0,1)$ | $1 - e^{-\lambda x}$, $x \ge 0$ | $\frac{-B}{d-1}$ | $\frac{1-y}{y(d-y)}\left[-\frac{A}{y} + \frac{d(d-1)/4}{d-y}\right] + \frac{u_0}{y} + \frac{u_1}{d-y} - u_0 - \frac{u_1}{d-1}$ | $D = u_0 + \frac{u_1 - B}{d-1} + \frac{1}{4}c^2$ <br> $E = u_0 d - B - \tilde{E}$ |

**Table V**

| $(a,b)$ | $y(x)$ | $2\mathcal{E}/\lambda^2$ | $2V(x)/\lambda^2$ $[u_i = 2V_i/\lambda^2]$ | Parameter Relations |
|---|---|---|---|---|
| $\left(\frac{1}{2},\frac{1}{2}\right)$ | $\frac{1}{2}[1+\sin(\lambda x)]$, $-\frac{\pi}{2} \le \lambda x \le +\frac{\pi}{2}$ | $\frac{1}{4}c^2 - D$ | $\frac{1}{d-y}\left[u_0 - \frac{A}{y} + \frac{B}{1-y}\right]$ | $E = u_0 - \tilde{E} + d(D - \frac{1}{4}c^2)$ |
| $\left(\frac{1}{2},1\right)$ | $\tanh^2(\lambda x/2)$, $x \ge 0$ | $\frac{-B}{d-1}$ | $\frac{1-y}{d-y}\left[u_1 y - \frac{A}{y} + \frac{u_0}{1-y}\right] - \frac{u_0}{d-1}$ | $E = \frac{B-u_0}{d-1} - \tilde{E}$ <br> $D = \frac{1}{4}c(c+1) - u_1$ |
| $(1,1)$ | $(1+e^{-\lambda x})^{-1}$, $-\infty \le \lambda x \le +\infty$ | $\frac{-B}{d-1}$ | $\frac{y(1-y)}{d-y}[u_2 + u_1 y] + \frac{u_0}{d-y} - \frac{u_0}{d-1}$ | $A = \frac{u_0 - Bd}{d-1}$ <br> $D = \frac{1}{4}c(c+2) - u_1$, $E = u_2 - \tilde{E}$ |
| $(0,1)$ | $1 - e^{-\lambda x}$, $x \ge 0$ | $\frac{-B}{d-1}$ | $\frac{u_0}{y} + \frac{u_1}{d-y} - \frac{A(1-y)}{y^2(d-y)} - u_0 - \frac{u_1}{d-1}$ | $D = u_0 + \frac{u_1 - B}{d-1} + \frac{1}{4}c^2$ <br> $E = u_0 d - B - \tilde{E}$ |

**Table VI**

| $(a,b)$ | $y(x)$ | $2\mathcal{E}/\lambda^2$ | $2V(x)/\lambda^2$ $[u_i = 2V_i/\lambda^2]$ | Parameter Relations |
|---|---|---|---|---|
| $\left(\tfrac{1}{2},\tfrac{1}{2}\right)$ | $\tfrac{1}{2}[1+\sin(\lambda x)]$, $-\tfrac{\pi}{2} \le \lambda x \le +\tfrac{\pi}{2}$ | $c^2/4$ | $-\dfrac{A/d}{y} + \dfrac{B/(d-1)}{1-y}$ | None |
| $\left(\tfrac{1}{2},1\right)$ | $\tanh^2(\lambda x/2)$, $x \ge 0$ | $\dfrac{-B}{d-1}$ | $-(1-y)\left[\dfrac{c}{4}(c+1) + \dfrac{A/d}{y}\right]$ | None |
| $(1,1)$ | $(1+e^{-\lambda x})^{-1}$, $-\infty \le \lambda x \le +\infty$ | $\dfrac{-B}{d-1}$ | $-(1-y)\left[u_0 + \dfrac{c}{4}(c+2)y\right]$ | $A = u_0 d - \dfrac{Bd}{d-1}$ |
| $(0,1)$ | $1-e^{-\lambda x}$, $x \ge 0$ | $\dfrac{-B}{d-1}$ | $u_0\left[\dfrac{1}{y}-1\right] + \dfrac{A}{d^2}\left[1+\dfrac{d-1}{y} - \dfrac{d}{y^2}\right]$ | $\dfrac{B}{d-1} - \dfrac{c^2}{4} = u_0 - \dfrac{A}{d^2}$ |

**Table VII**

| $(a,b)$ | $x$ | $2V(x)/\lambda^2$ $[u_i = 2V_i/\lambda^2]$ | Parameter Relations |
|---|---|---|---|
| $\left(\tfrac{1}{2},\tfrac{1}{2}\right)$ | $-\tfrac{\pi}{2} \le \lambda x \le +\tfrac{\pi}{2}$ | $2\dfrac{u_- + u_+\sin(\lambda x)}{\cos^2(\lambda x)}$ | $u_\pm = \dfrac{B}{d-1} \pm \dfrac{A}{d}$ |
| $\left(\tfrac{1}{2},1\right)$ | $x \ge 0$ | $-\dfrac{A/d}{\sinh^2(\lambda x/2)} - \dfrac{c(c+1)/4}{\cosh^2(\lambda x/2)}$ | None |
| $(1,1)$ | $(1+e^{-\lambda x})^{-1}$, $-\infty \le \lambda x \le +\infty$ | $\dfrac{-1}{e^{\lambda x}+1}\left[u_0 + \dfrac{c(c+2)/4}{1+e^{-\lambda x}}\right]$ | $A = u_0 d - \dfrac{Bd}{d-1}$ |
| $(0,1)$ | $1-e^{-\lambda x}$, $x \ge 0$ | $\dfrac{u_0}{e^{\lambda x}-1} - \dfrac{A}{2d^2}\dfrac{d+1-e^{-\lambda x}}{\cosh(\lambda x)-1}$ | $\dfrac{B}{d-1} - \dfrac{c^2}{4} = u_0 - \dfrac{A}{d^2}$ |